\documentclass[12pt]{iopart}

\usepackage{graphicx} 
\usepackage{subfiles}
\usepackage{caption}
\newcommand\captionof[1]{\def\@captype{#1}\caption}

\usepackage{iopams}
\usepackage{xcolor}
\usepackage{mwe}
\usepackage{subfig}
\usepackage[greek,english]{babel}

\begin{document}

\title[xeSFQ: Clockless SFQ Logic with Zero Static Power]{xeSFQ: Clockless SFQ Logic with Zero Static Power}

\author{J. Volk, G. Tzimpragos, O. Mukhanov}

\address{University of Wisconsin-Madison; SEEQC}
\ead{jennifer.volk@wisc.edu}
\vspace{10pt}
\begin{indented}
\item[]November 2024
\end{indented}

\begin{abstract}
ERSFQ circuits eliminate the dominant portion of static power consumption in RSFQ circuits by using current-limiting Josephson junctions and inductors instead of bias resistors. In practice, these junctions still contribute to static power consumption through switching required to correct phase imbalances across the circuit, with their contributions sometimes comparable to dynamic power. This paper presents a new SFQ family variant, called xeSFQ, that combines the clock-free alternating SFQ logic with ERSFQ's biasing. By ensuring a single pulse per line per logical cycle, xeSFQ eliminates even the residual switching due to phase imbalance, achieving truly zero static power consumption. Detailed analog simulations and synthesis results for various circuits, from single gates to ISCAS85 and EPFL benchmarks, validate the above hypothesis and showcase the all-around benefits of the proposed approach.

\end{abstract}


\section{Introduction}\label{sec:intro}
Modern superconductor electronics, based on the exchange of single flux quanta (SFQ), have emerged over the past two decades as a promising candidate for energy-efficient computing~\cite{superrl, largescale}, in addition to their original promise of ultra-high speeds~\cite{770ghz}. In this respect, researchers have made contributions on multiple fronts, from optimizing dynamic and static power consumption to improving cryocooler technology. While cryocooler advancements are a significant part of the overall picture~\cite{radebaugh2022review}, they fall outside the scope of this paper, which focuses on SFQ circuit design. Specifically, the aim is to eliminate static power consumption while simultaneously enhancing circuit simplicity and reducing dynamic power consumption.

Static power in SFQ circuits is primarily associated with biasing~\cite{histaticpwr}. Optimizing biasing is not a new area of research. In the original Rapid SFQ (RSFQ) technology~\cite{rsfq}, biasing is achieved through a network of resistors. These resistors distribute bias current to the Josephson junctions (JJs) in the logic gates, providing the necessary directionality to control switching. However, resistors account for a constant power drain, comprising more than 90\% of the total power consumption according to the literature~\cite{histaticpwr}. This spurred the search for alternatives, leading to the development of two energy-efficient RSFQ variants: ERSFQ and eSFQ, introduced about a decade ago~\cite{ersfqesfq}. The key innovation in these variants is replacing the resistor-based biasing network with one composed of JJs and inductors, which eliminates the power-hungry resistors and led to the belief that static power dissipation could be reduced to zero---marking a significant advance in the field.

In the ideal case, when each bias injection point is modeled with an independent current source, this claim holds true. However, when the bias circuitry is reflected with bias inductors and JJs connected to a common line~\cite{ersfq_currentredistribution}, current redistribution in the bias network occurs. This means that while eSFQ achieves zero static power consumption, the same is not necessarily true for ERSFQ in practice. In ERSFQ, static power is significantly reduced but not entirely eliminated due to phase accumulation across different parts of the circuit. This, in turn, results in bias JJs that switch~\cite{switchingBiasJJs_ERSFQ}. Our findings indicate that the static power consumed during bias JJ switching is approximately 16\% of the dynamic power in individual gates and reaches 72\% of the dynamic power in larger designs. Meanwhile, although eSFQ avoids this issue, its design---specifically, the placement of bias injection points at the clock ports of each gate---ties biasing to gate-level clocking. As discussed in prior research, the reliance on a clock signal at this granularity and delivered in this manner presents significant challenges, including ad hoc tooling support~\cite{supertools}, increased circuit complexity~\cite{fakeAsync,dffinsertion,ersfqesfq}, and distinct architectural constraints~\cite{optpipeline,jimpipeline,xsfq}, making its practical implementation rather limited.

In this paper, we introduce \textit{xeSFQ}, a new SFQ family that achieves truly zero static power dissipation without the clocking-biasing coupling seen in eSFQ. In fact, its logic gates rely on mature, clock-free cells that require neither explicit reset signals nor specialized reset circuitry~\cite{dsfq,pcl,rql}, and do not depend on custom logic synthesizers, paving the way for broader advancements and increased efficiency across the board~\cite{volk2024synthesis}. Specifically, xeSFQ is the result of combining the abstractions of Alternating SFQ (xSFQ)~\cite{xsfq} and ERSFQ’s biasing approach. The key distinction from ERSFQ, in terms of static power, lies in xSFQ's alternating encoding, which guarantees a single pulse per line per computational cycle and allows for fully symmetric switching. This is crucial because, as we formalize in the rest of the paper, there is a direct relationship between data values (represented as pulse sequences), phase accumulation, and the switching of JJs in the bias network.

To verify our hypotheses, we conduct a series of detailed analog simulations, complemented by synthesis results for larger designs. First, we demonstrate xeSFQ’s zero static power consumption at the gate level, comparing it to that of ERSFQ. Next, we establish composability by showcasing the same elimination of static power consumption in a full-adder circuit. Finally, we present JJ count and energy results from synthesized ISCAS85~\cite{iscas85} and EPFL~\cite{epfl} benchmark circuits.

\section{Background}\label{sec:background}
\subsection{DC biasing mechanisms}
The critical current ($I_C$) is the defining characteristic of a JJ, marking the threshold between its two operational states: one where the voltage across the junction is zero, and another where a nonzero voltage appears. Each JJ is typically biased at 70\% of its $I_C$ to enable proper circuit operation. In traditional RSFQ designs, this biasing is done through a resistive network. The resistors connect a voltage rail (typically 2.6 mV) to the JJs for biasing. These JJs are grounded on one side~\cite{rsfq} and, in the steady state, act as superconducting shorts. The result is a constant flow of current through the bias resistors, leading to static power consumption of hundreds to thousands of nanowatts per gate~\cite{ersfqesfq}.

Energy-efficient alternatives like ERSFQ and eSFQ use a different DC biasing strategy~\cite{ersfqesfq}. Instead of resistors, they employ a network of ``bias'' JJs and inductors to regulate and distribute current. These bias JJs are designed with $I_C$ values that match the current otherwise provided by equivalent bias resistors, while bias inductors are typically sized in the hundreds of picohenries to suppress unwanted transients~\cite{ersfqesfq,biasind200pH}. Bias JJs act as current limiters, switching only when the current exceeds their $I_C$. When ideal, independent bias sources are simulated at each circuit bias port, they remain in the superconducting state without switching, theoretically reducing static power consumption to zero. However, in the more realistic scenario, in which the inductor-JJ biasing network is accurately represented and all bias ports are connected through a common line~\cite{ersfq_currentredistribution}, current redistribution occurs, leading to phase accumulation across the circuit and bias JJ switching, as will be discussed in Section~\ref{sec:xesfq}.

\subsection{Phase accumulation \& current redistribution}\label{sec:imbalance}
Each time the current through a JJ exceeds its $I_C$, the JJ switches and its phase parameter increases by a fixed amount of $+2\pi$. The phase of a JJ is an internal quantum property related to the superconducting order parameter~\cite{phasedef}. A JJ's phase change corresponds to the generation of an SFQ; the SFQ's magnetic nature causes current to flow as it travels through the superconductor~\cite{ersfq_currentredistribution,tinkham}. 

\begin{figure}
    \centering
    \includegraphics[width=0.28\linewidth]{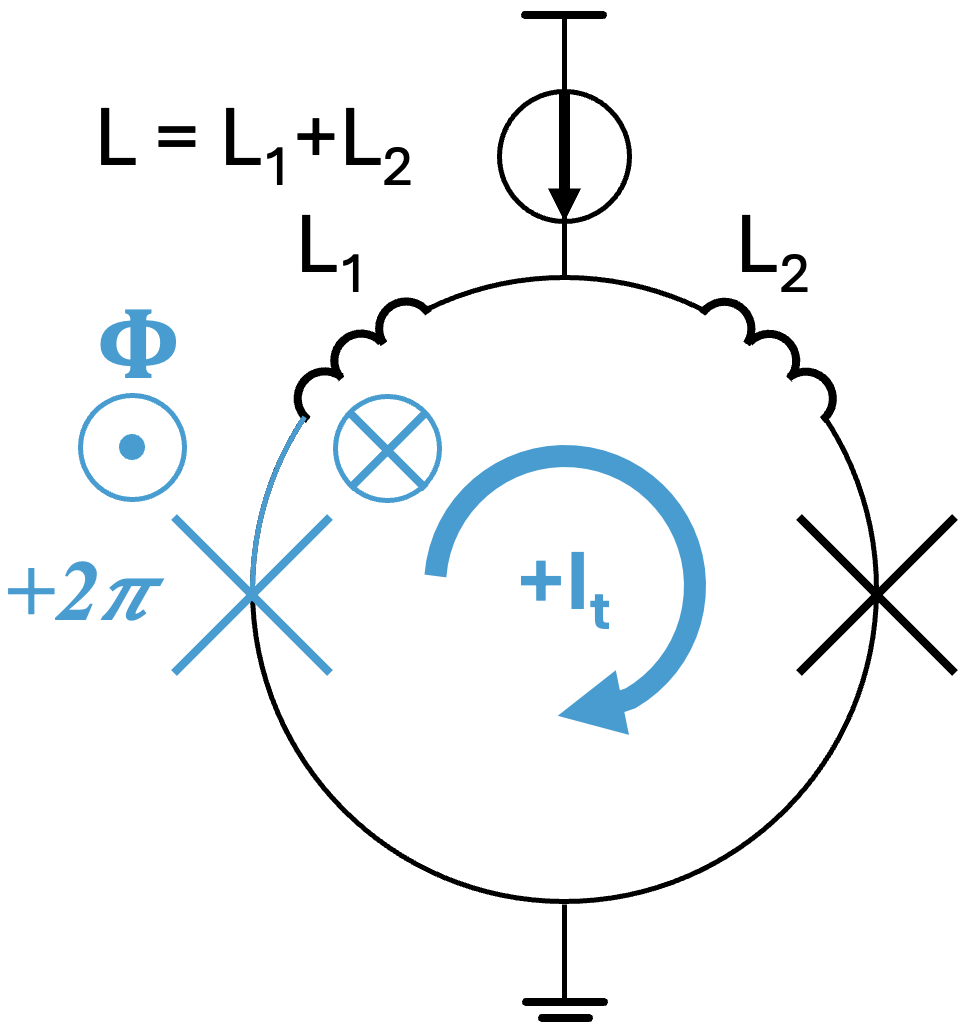}
    \caption{A DC SQUID with inductance L. When the left JJ switches, its phase increases by $2\pi$ and causes a clockwise change in transport current of $+I_t$.}
    \label{fig:squid}
\end{figure}

The relationship between flux and current flow in superconductor electronics is best exemplified in a DC Superconducting QUantum Interference Device (SQUID)~\cite{rsfq}, which consists of two JJs and a large inductor (larger than its quantizing inductance\footnote{Quantizing inductance is defined by the following equation, for which the quantizing parameter $\textgreek{β}_L \geq 1$: $L = \frac{\textgreek{β}_L}{I_C \Phi_0}$, where $\Phi_0$ is equal to the magnetic flux quantum, 2.07$\times$10$^{-15}$ Wb.}).
Assume that in the SQUID design shown in Figure~\ref{fig:squid}, where both JJs start with zero phase, the left JJ switches once. 
The switch causes the left JJ to gain a phase of $2\pi$, which is sustained by the inductor. The introduction of this flux also causes a current to flow clockwise towards the lower-phase node.

\noindent

The use of DC SQUIDs is essential for the realization of SFQ logic gates as they enable statefulness through the storage of SFQ. Typically, one SFQ is stored per SQUID in such gates. However, with a large enough inductor---a circuit structure appearing frequently in the ERSFQ biasing network---the SQUID can host higher-order phases. In this case, each additional $+2\pi$ increment adds to the phase parameter at that node, contributing to a process aptly named \textit{phase accumulation}. Phase accumulation causes additional transport currents to be steered toward the lower-phase node. The difference in phases between the two nodes is coined \textit{phase imbalance}, a term which we will use to refer to this process throughout the rest of this manuscript. The equation mapping phase accumulation to transport current in a DC SQUID with inductance $L$ follows:

\begin{equation}
\label{faraday}
    (\frac{\Phi_0}{2\pi}) \frac{d\phi}{dt} = L \frac{dI_t}{dt}.
\end{equation}

\noindent This reduces trivially to $\Phi_0 = L I_t$, which can be used to calculate the step-wise increase in current $I_t$ for each corresponding $+2\pi$ phase increase.

\subsection{Data representation}
\begin{figure}[]
    \centering
    \hspace{0.01cm}
    \subfloat[\label{subfig:rsfqenc}]{%
    \includegraphics[width=0.18\textwidth]{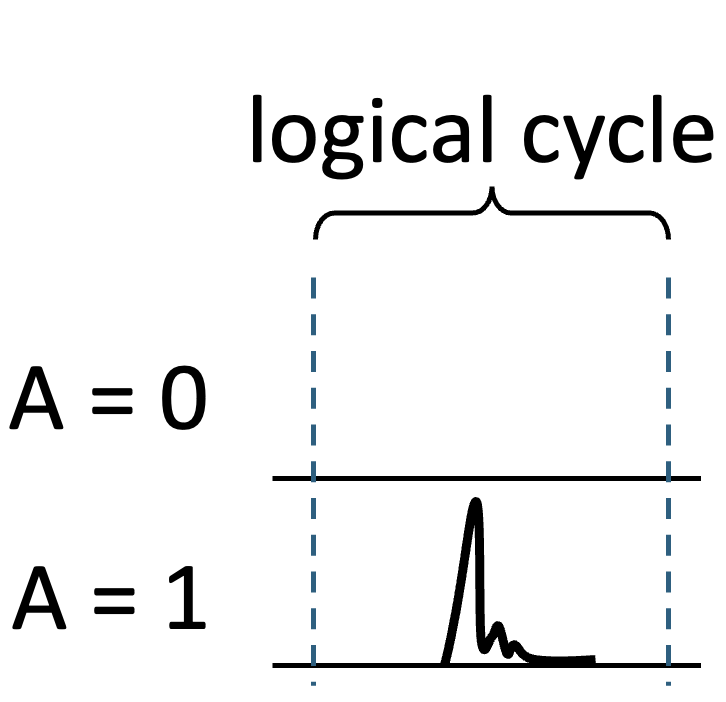}
    }
    \hspace{0.62cm}
    \subfloat[\label{subfig:xsfq}]{%
      \includegraphics[width=0.62\textwidth]{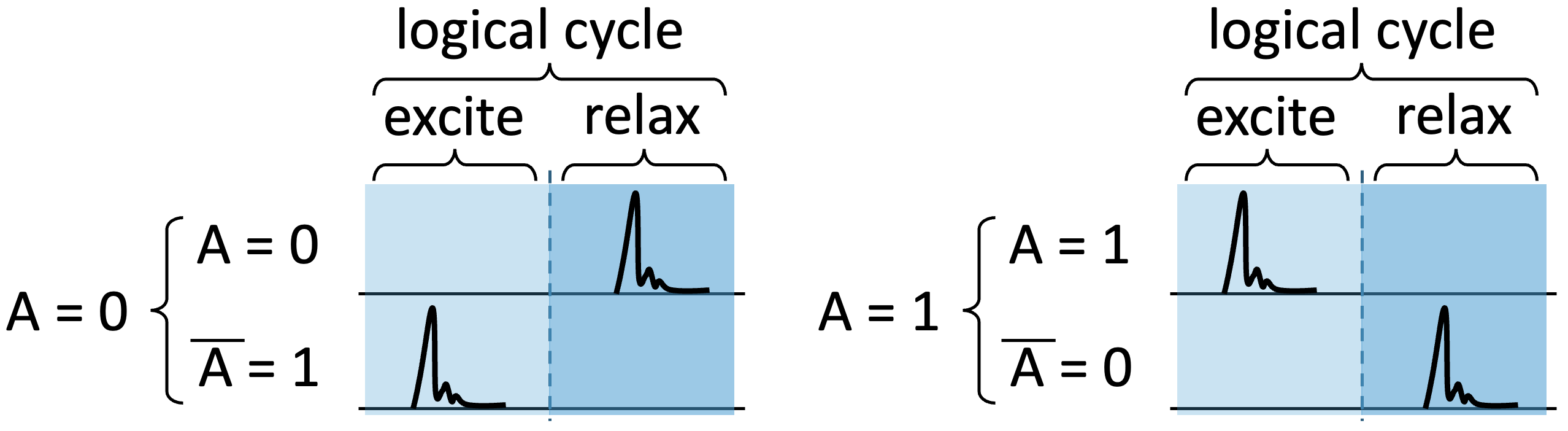}
    }
    \caption{Panel a: Direct encoding, where the presence of a pulse during the designated interval represents a logical 1, and its absence represents a logical 0. Panel b (left): Alternating, balanced encoding representing a logical 0. The logical cycle consists of two synchronous cycles: excite, where the data value is presented, and relax, where the complement is presented. Panel b (right): Alternating, balanced encoding representing a logical 1.}
    \label{fig:enc}
  \end{figure}

Data encoding varies significantly across different SFQ logic families, particularly between ERSFQ and xSFQ, with the latter serving as the foundation for the proposed energy-efficient logic family. The standard encoding method used in ERSFQ (as well as RSFQ and eSFQ)~\cite{ersfqesfq} is illustrated in Figure \ref{subfig:rsfqenc}. A logical 1 is represented by the presence of a pulse on a wire during the corresponding clock cycle, while its absence indicates a logical 0.


In contrast, xSFQ~\cite{xsfq} employs an alternating, balanced encoding scheme. In this case, a pulse appears on each wire during every logical cycle, regardless of the corresponding logical value (0 or 1). To differentiate between these values, each logical cycle is divided into two synchronous cycles: excite and relax. During the excite cycle, the data value is presented (and computation occurs); in the relax cycle, the complement is presented (allowing the logic gate to return to its initial state, ready for the next computation round). According to this scheme, a logical 0 is encoded by the absence of a pulse during the excite cycle, followed by a pulse during the relax cycle (Figure \ref{subfig:xsfq}, left), while the opposite occurs for a logical 1 (Figure \ref{subfig:xsfq}, right).

\section{Proposed xeSFQ Logic Family}\label{sec:xesfq}
In this section, we first analyze the static power consumption in ERSFQ, which, according to our findings, is not zero when the inductor-JJ biasing network is appropriately modeled. We then introduce an energy-efficient variant of xSFQ, xeSFQ, which leverages its balanced encoding to achieve truly zero static power consumption. 

\subsection{Static power consumption in ERSFQ}
\begin{figure}[]
    \hspace{0.01cm}
    \subfloat[\label{subfig:syncandgate}]{%
    \centering
    \includegraphics[width=0.45\textwidth]{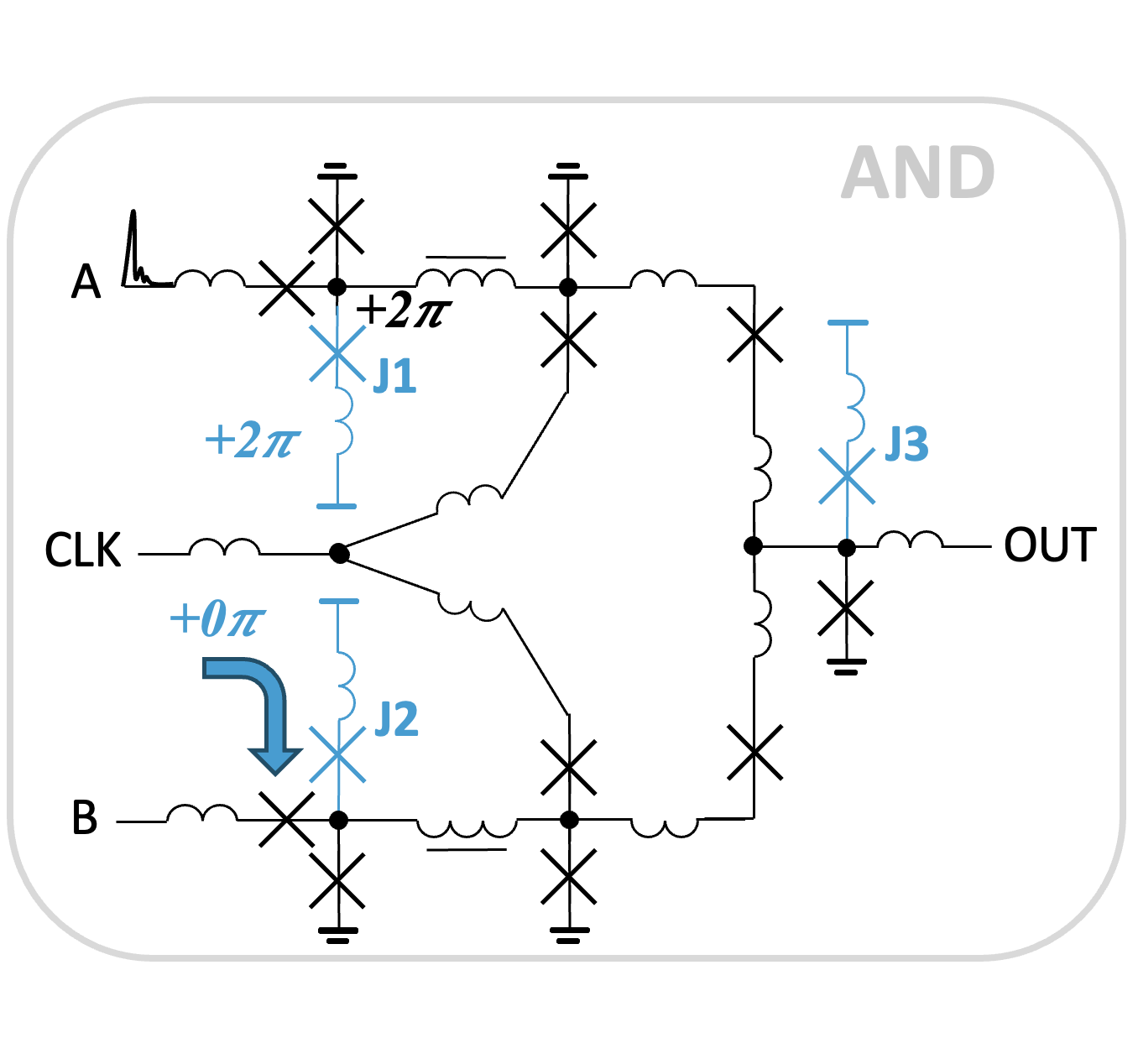}
    }
    \hspace{0.5cm}
    \subfloat[\label{subfig:syncandsim}]{%
      \includegraphics[width=0.5\textwidth]{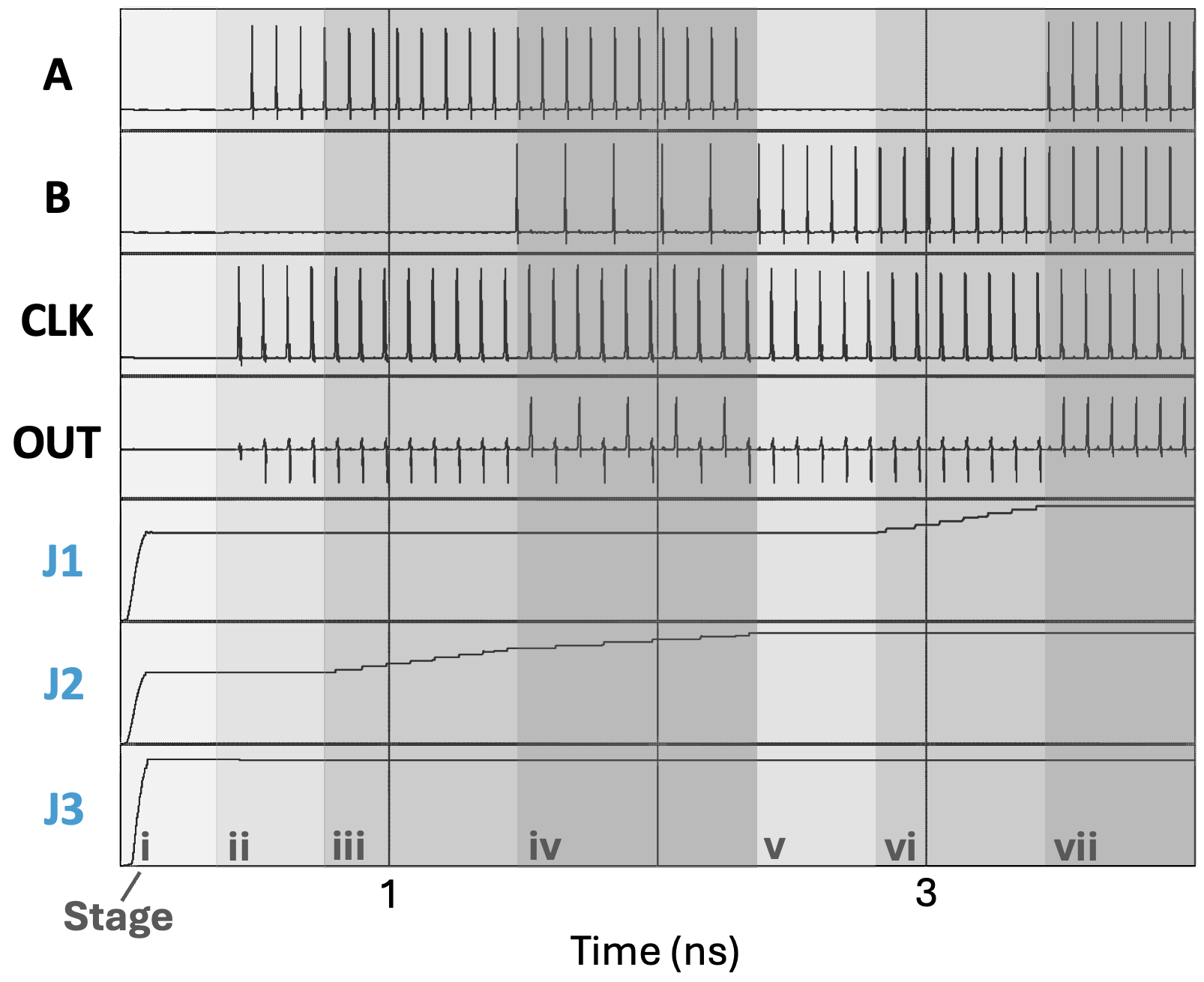}
    }
    \caption{Panel a: Synchronous AND gate with ERSFQ biasing. Large data imbalances across the circuit lead to phase accumulation on one side of the circuit, which trickles up to the bias line and causes additional current to flow into bias ports at lower phases. Panel b: Simulation results include the input SFQ signals on the A, B, CLK, and output ports, as well as the bias JJ phases. Stage i: Bias current ramps up until bias JJs begin switching. Bias is then lowered. Stage ii: Back-to-back SFQ on the A input build up a phase imbalance. Stage iii: Continued SFQ on A (and no SFQ on B) cause current through J2 to increase. J2 begins switching, once for every SFQ on A. Stage iv: Continued back-to-back SFQ on A with moderate switching on B reduces the J2 switching rate to once every other SFQ on A. Stage v: Back-to-back SFQ on B (and no SFQ on A) equalize the phase difference; J2 stops switching. Stage vi: More SFQ on B reverse the phase imbalance and cause current through J1 to increase; J1 switches once for every SFQ on B. Stage vii: Balanced data rate on A and B alleviates additional phase accumulation; J1 stops switching.}
    \label{fig:syncand}
  \end{figure}

As discussed in Section~\ref{sec:background}, phase accumulation and phase imbalance are directly related to transport current in circuit structures that resemble DC SQUIDs. Under ERSFQ's direct encoding, phase imbalance arises as the circuit's JJs switch asymmetrically across the design. For example, a JJ receiving a stream of 1s, corresponding to a stream of SFQ, will have a higher phase than another JJ that receives a stream featuring fewer or no 1s. This can be noticed in the SPICE-level simulations of an ERSFQ \textsc{AND} gate design (Figure~\ref{fig:syncand}a).

The design's baseline $I_C$ is 250~$\mu$A, its bias inductors are 200~pH, and its three bias JJs have $I_C$s J1=J2=125~$\mu$A and J3=250~$\mu$A. In Figure~\ref{fig:syncand}b, after the biasing startup sequence (Stage i), the current through both J1 and J2 is 100~$\mu$A each. A stream of SFQ is then presented on A. After each SFQ on A, the top left JJ switches, incurring a phase increase of $+2\pi$ each time. The clock also introduces an additional $+2\pi$ phase increase with each logical cycle. This combined $+4\pi$ phase passes through J1 and develops a phase imbalance hotspot, steering roughly 6~$\mu$A of current per SFQ into J2 (Stage ii). After enough logical cycles elapse (4 in this case, which is in line with calculations), J2 reaches its critical current and begins switching in response, once for each additional SFQ on A (Stage iii)~\footnote{Note that the maximum number of logical cycles that can pass before the onset of phase imbalance decreases with the bias inductor and JJ sizes.}. When SFQ are introduced on B, the switching rate slows (Stage iv) and halts when the SFQ stream on A stops (Stage v). The phase accumulation equalizes and then changes sides after enough SFQ pass on the B and clock lines. At this point, J1 reaches its critical current and switches in response (Stage vi). Finally, J1 and J2 stop switching when SFQ arrive on both A and B at equal rates (Stage vii). 

Each time J1 and J2 switch, an amount of power $P_S$ equal to

\begin{equation}
    P_{S} = \Phi_0 I_C f
\end{equation}

\noindent is consumed, where $f$ is the switching rate. In other words, the switching of the bias JJs indicate a non-zero static power consumption. In the case where static power is consumed (either on A or on B), it is equivalent to 15.9\% of the dynamic power per logical cycle. The effects of phase imbalance, and this percentage, increases in ERSFQ circuits with more gates, as will be discussed in Section~\ref{sec:eval}.

\subsection{xeSFQ}\label{subsec:xesfq}
We propose xeSFQ, a new SFQ logic family that achieves truly zero static power dissipation by preventing phase accumulation and imbalance in the datapath, thereby eliminating the switching of the bias JJs. To accomplish this, xeSFQ combines xSFQ's balanced encoding with ERSFQ biasing, all while providing simple, metastability-free, and compact primitive cells. 

\begin{figure}[]
    \centering
    \hspace{0.01cm}
    \subfloat[\label{subfig:mealyC}]{%
    \includegraphics[width=0.35\textwidth]{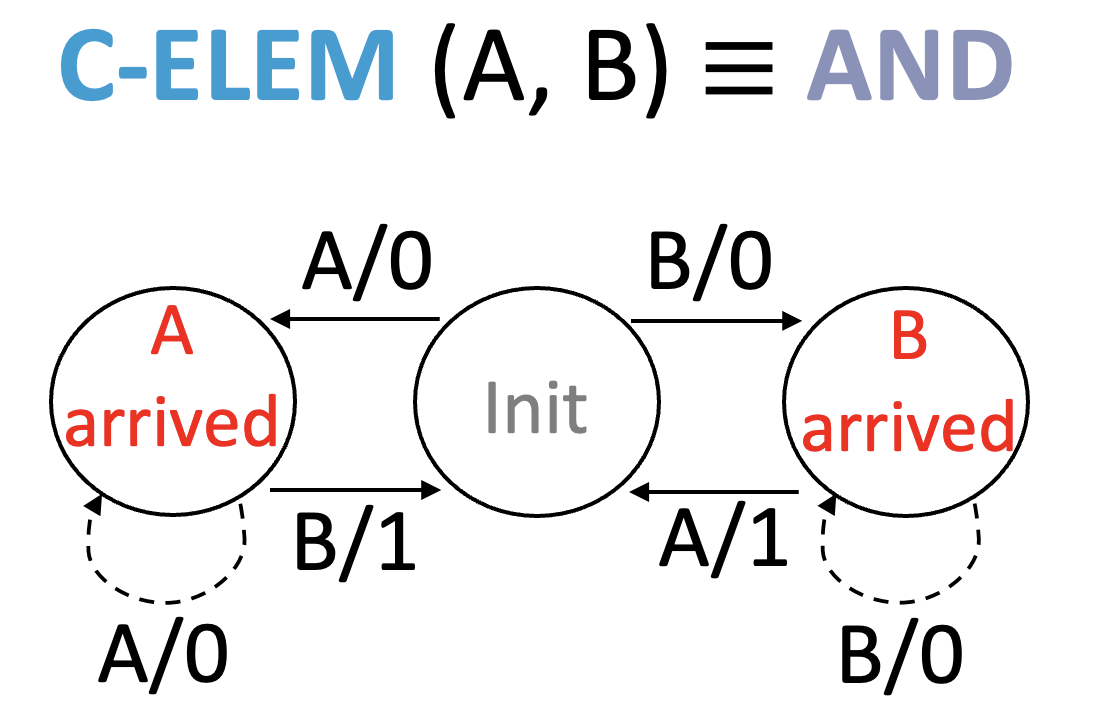}
    }
    \hspace{0.75cm}
    \subfloat[\label{subfig:mealyIC}]{%
      \includegraphics[width=0.35\textwidth]{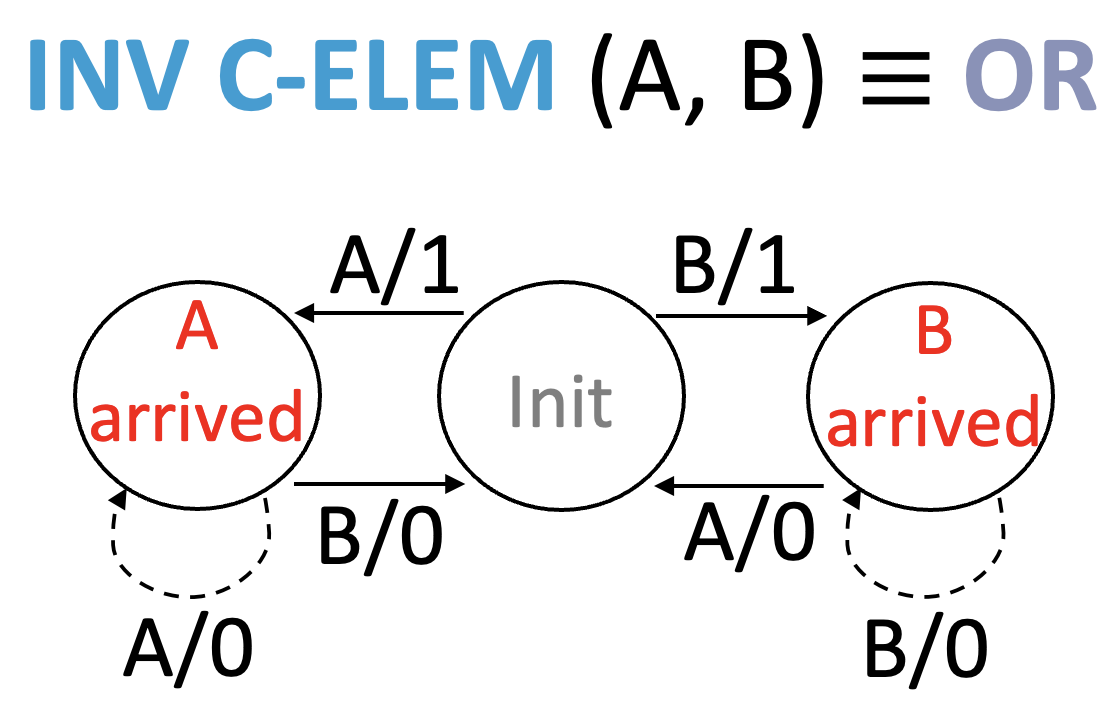}
    }
    \caption{Mealy machine descriptions of the C- and Inverted C-elements. Panel a: The C-element outputs a 1 when pulses arrive on both inputs, signaling a return to its \textit{Initial} state. Panel b: The Inverted C-element outputs a 1 as soon as a pulse arrives on either input, indicating a departure from its \textit{Initial} state. Under x(e)SFQ, these two elements function as \textsc{AND} and \textsc{OR} gates, respectively, without the need of an external clock signal for evaluation and resetting, as required in typical RSFQ logic variants.}
    \label{fig:mealy}
  \end{figure}

Similar to xSFQ, xeSFQ's two primary operators \textsc{Last Arrival} (LA) and \textsc{First Arrival} (FA) are implemented as C- and Inverted C-elements. Following their Mealy machine description, Figure~\ref{fig:mealy}, these elements operate as follows: the C-element outputs a 1 only when pulses arrive on both inputs, signaling a return to its Initial state, while the Inverted C-element outputs a 1 upon the receipt of just one input, indicating a departure from its \textit{Initial} state. As discussed in prior work~\cite{xsfq, volk2024synthesis}, these elements function as \textsc{AND} and \textsc{OR} gates, evaluate logically without a clock signal, and are reset through alternating signaling using a balanced code. During the excite cycle, the actual data value is represented by the presence or absence of a pulse, with its complement appearing during the relax cycle. This alternating signaling approach enables the relax cycle to serve as a resetting mechanism while ensuring each JJ receives exactly one SFQ per logical cycle.

\begin{figure}[]
    \centering
    \hspace{0.01cm}
    \subfloat[\label{subfig:schema}]{%
    \includegraphics[width=0.32\textwidth]{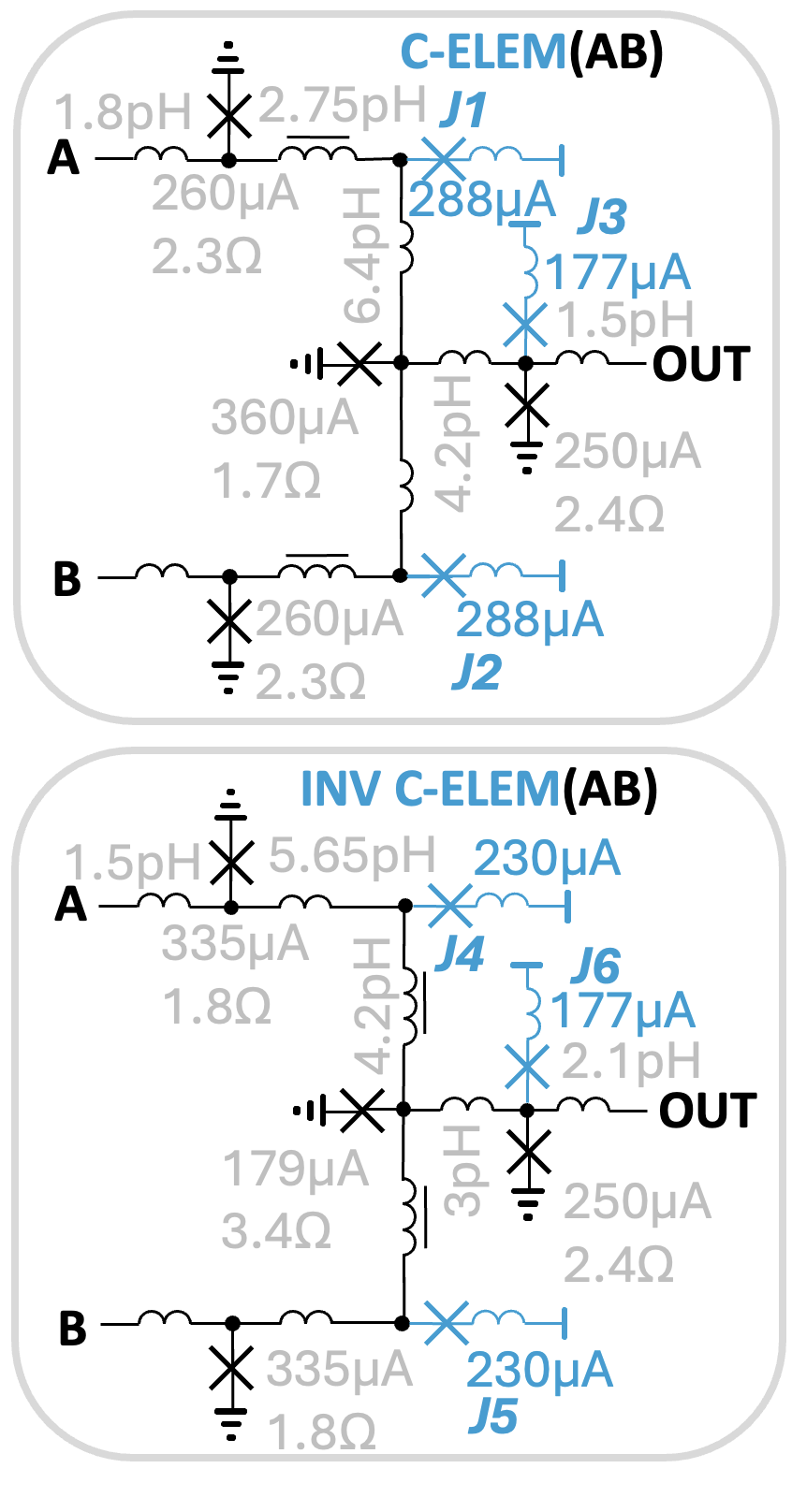}
    }
    \hspace{0.75cm}
    \subfloat[\label{subfig:wvfm}]{%
      \includegraphics[width=0.51\textwidth]{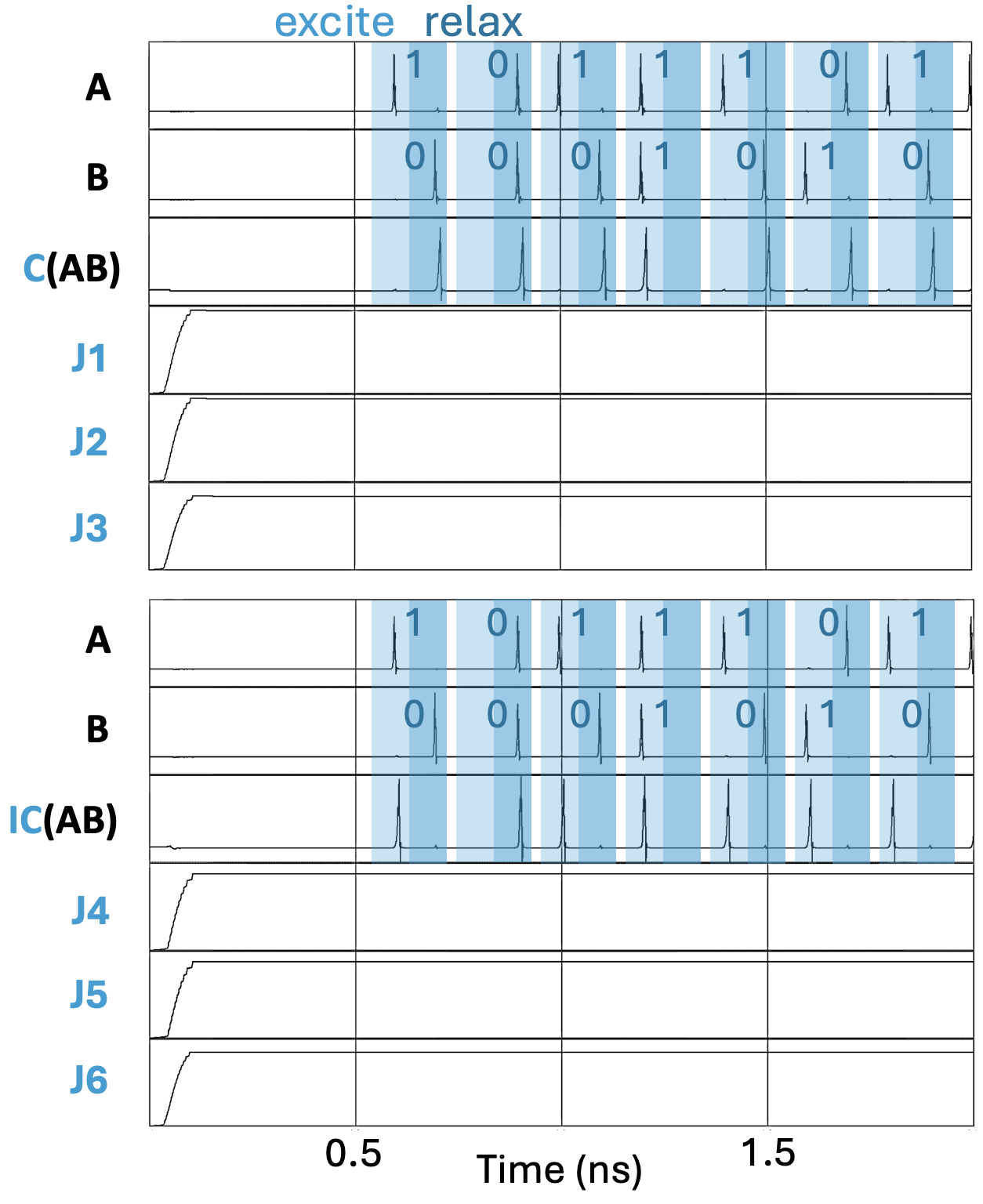}
    }
    \caption{Panel a: Schematics for the C- and Inverted C-element circuits. The components highlighted in blue represent the biasing network, with bias inductors selected to be 200~pH. Panel b: Simulation waveform illustrating the functional behavior of these gates using data streams (A = 7'b1011101, B = 7'b0001010) with x(e)SFQ's balanced, alternating encoding. The bias JJ phases for both elements remain constant during operation, indicating zero phase imbalance and, consequently, zero static power consumption. This contrasts with the synchronous \textsc{AND} gate shown in Figure~\ref{fig:syncand}a, where unbalanced data encoding results in nonzero static power consumption.}
    \label{fig:schema}
  \end{figure}

Figure~\ref{fig:schema}a shows the circuit schematics for the proposed xeSFQ C- and Inverted C-elements. SFQ are stored in the SQUIDs, represented by 3-loop inductors with straight lines. In the C-element, the input SFQ switch the input JJs and are stored in their adjacent SQUIDs. The central JJ has a higher critical current, requiring two SFQ to switch; once this occurs, the rightmost JJ switches and generates a pulse at the output. In the Inverted C-element, the central JJ switches after just one SFQ arrives (e.g., from input A), as its critical current is significantly lower than that of the input JJs. The SFQ generated by this JJ splits between the output and the backwards storage loop on the opposite input's line (B), where it will remain stored until that input (B) arrives. 
In both gates, every JJ experiences an equal phase increase of $+2\pi$. This behavior is illustrated in Figure~\ref{fig:schema}b, in which a balanced input sequence---independent of the A and B values---prevents the bias JJ phases from increasing.

\section{Evaluation}
The analog simulations presented in Section~\ref{sec:xesfq} highlight the difference in phase accumulation---and, consequently, static power consumption---between ERSFQ (Figure~\ref{fig:syncand}) and xeSFQ circuits (Figure~\ref{fig:schema}) at the gate level. In the former, direct data encoding, with its uneven number of pulses, led to phase imbalance, first in J2 and then in J1, with the slope depending on the inequality of 1s between the A and B signals. In the latter, however, the bias JJ phases remained unchanged, regardless of the sequence of logical 0s and 1s. This section aims to demonstrate that this behavior holds true for xeSFQ designs incorporating multiple gates, confirming their composability. All simulations are performed using WRSPICE~\cite{wrspice} with the resistively and capacitively shunted junction (RCSJ) model for JJs in the MITLL SFQ5ee 10~kA/$\mu$m$^2$ process~\cite{phasedef,sfq5ee}. 

\subsection{Bias inductor and current tuning}
As with any circuit parameter, selecting the appropriate bias inductor size and current level is important for reliable operation. With respect to the first, bias inductors distribute current to individual bias ports from a common bias line while smoothing out transients and maintaining bias current directionality. Consequently, they are the first line of defense against bias JJ switching in the case of phase imbalance. 

Larger inductance values result in smaller transport currents (Equation~\ref{faraday}) and are typically sized in the hundreds of picohenries (e.g., 200-400~pH). In layout, this makes them relatively large components (tens of $\mu$m$^2$ per inductor, even when using high-kinetic inductance materials~\cite{sfq5ee}). However, because xeSFQ designs eliminate the risk of phase imbalance, they allow for bias inductors about an order of magnitude smaller (e.g., 60~pH in the designs presented here) while still maintaining zero static power consumption. Further reductions in bias inductance may compromise transient suppression, potentially causing bias JJ switching between excite and relax cycles.

For bias current values at or above the sum of all bias JJ critical currents, the bias JJs will switch, resulting in marginal but non-zero static power dissipation, regardless of the underlying logic scheme. However, when the bias current remains below this threshold, no such switching occurs. Our simulation results show that the proposed C- and Inverted C-elements operate reliably with bias currents down to 40.0\% and 19.8\% below their summed $I_C$ values, without any malfunction or static power consumption. The circuit parameters of these elements are adapted from prior work~\cite{volk2024synthesis,xsfq} and their JJs manually tuned to achieve desirable operation down to these bias levels. Further improvements may be possible through additional circuit optimizations and the use of feeding Josephson transmission lines~\cite{ersfqesfq,fjtl,fjtl_margins}, although such approaches are beyond the scope of this work.


In these simulations, a 500~ps rise time is used to ramp the current up to the sum of the bias JJ critical currents, followed by a 200~ps hold and a 300~ps fall time to bring the bias current down to the chosen point. This sequence and timing allow for the inductors to adapt and distribute bias current accordingly. If the rise time is too fast, excess current can leak undesirably into nearby circuit nodes, potentially causing JJs on the datapath to switch prematurely, leading to circuit errors. Similarly, the bias current is held for some time to allow distribution to stabilize before slowly ramping down to the target operating point. 

\begin{figure}
    \centering
    \includegraphics[width=0.35\paperwidth]{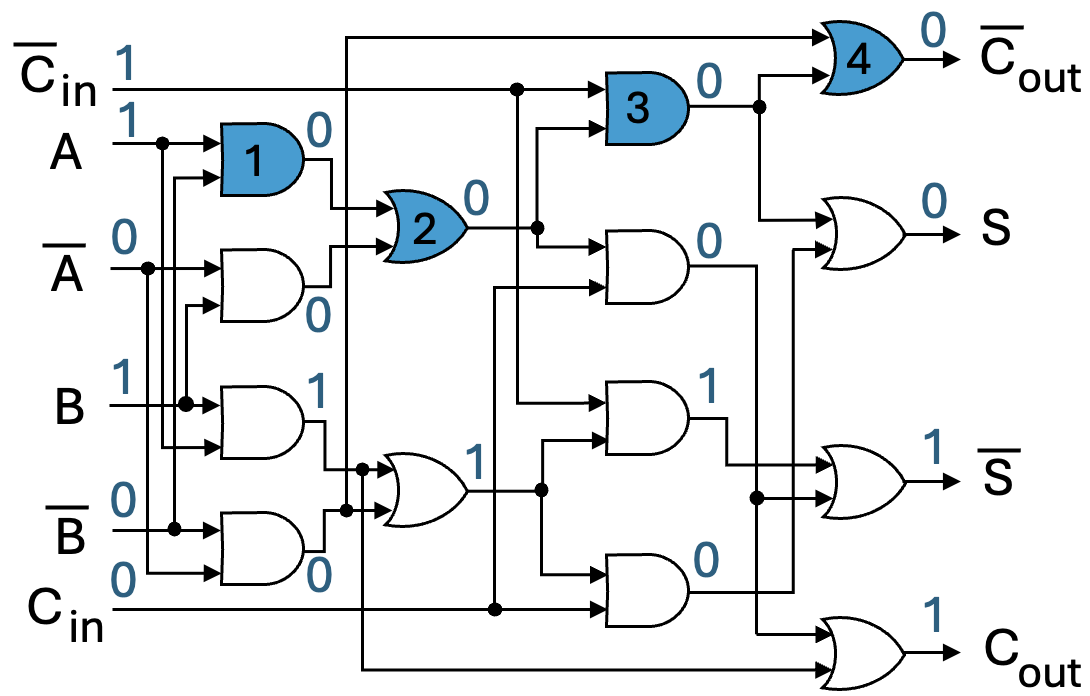}
    \captionof{figure}{An xeSFQ full adder implemented with C- and Inverted C-elements (\textsc{C} and \textsc{IC}, respectively) functioning as \textsc{AND} and \textsc{OR} gates, respectively. The example shown is for inputs $A$=1, $B$=1, and $C_{in}$=0, with the values propagating through the design. The resulting outputs are $C_{out}$=1 and $S$=0, as expected.}
    \label{fig:fulladder_block}
\end{figure}

\begin{figure}
    \centering
    \includegraphics[width=0.6\paperwidth]{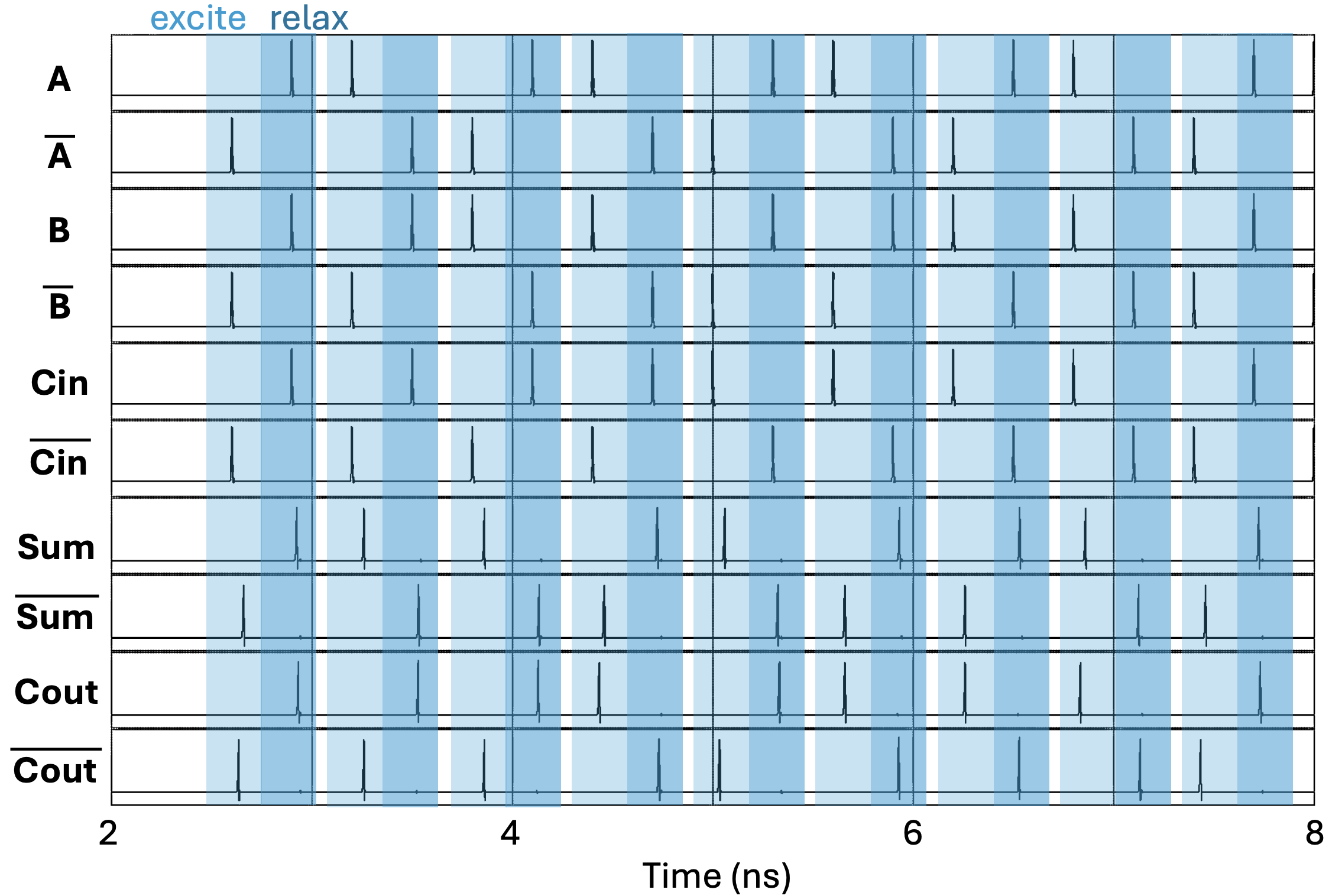}
    \captionof{figure}{Simulation waveform demonstrating the correct functionality of the xeSFQ full adder from Figure~\ref{fig:fulladder_block} across all possible input combinations.}
    \label{fig:fulladder_wvfm}
\end{figure}

\begin{figure}[]
\centering
    \hspace{0.01cm}
    \subfloat[\label{subfig:celem11}]{%
    \includegraphics[width=0.35\textwidth]{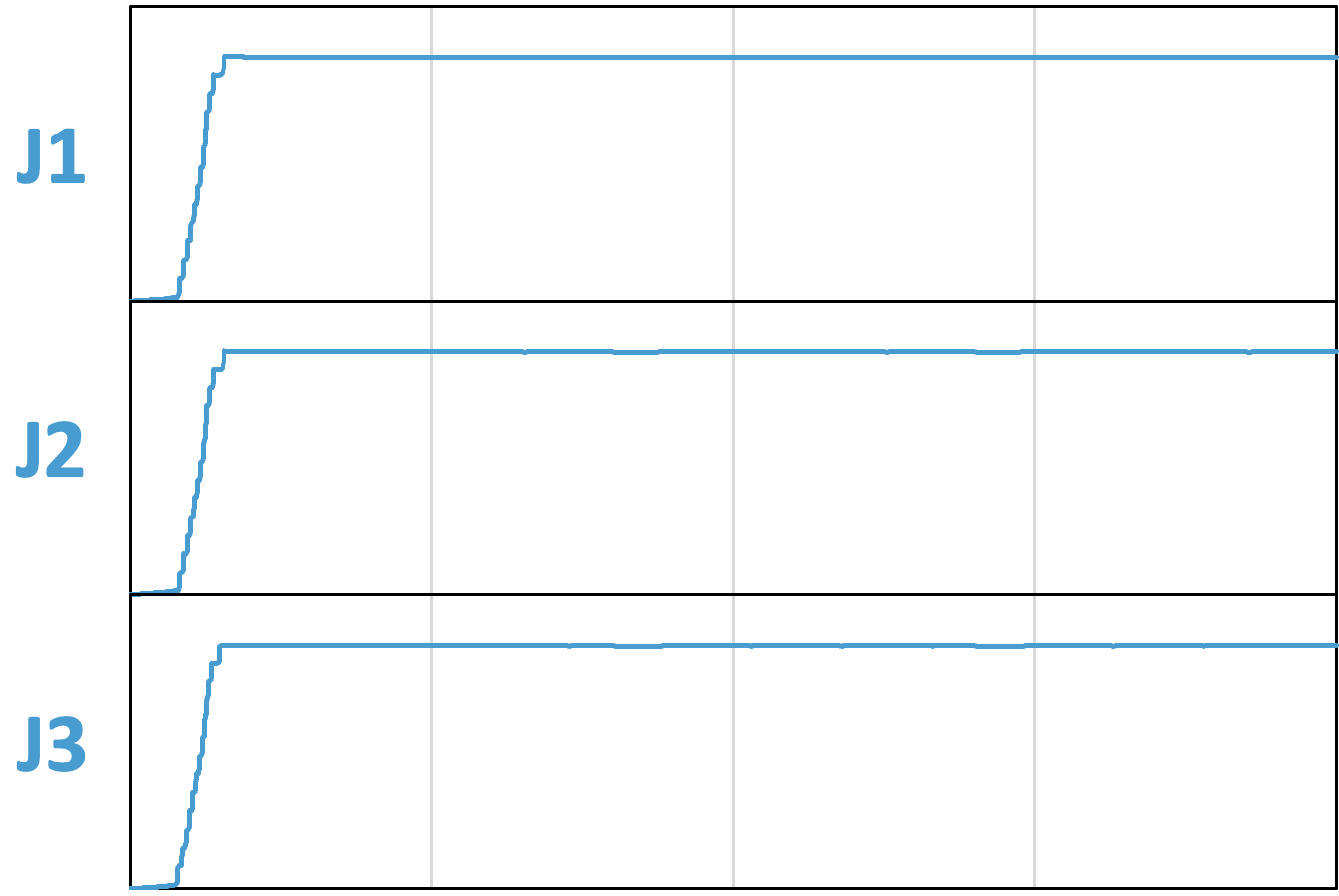}
    }
    \hspace{0.75cm}
    \subfloat[\label{subfig:icelem21}]{%
      \includegraphics[width=0.35\textwidth]{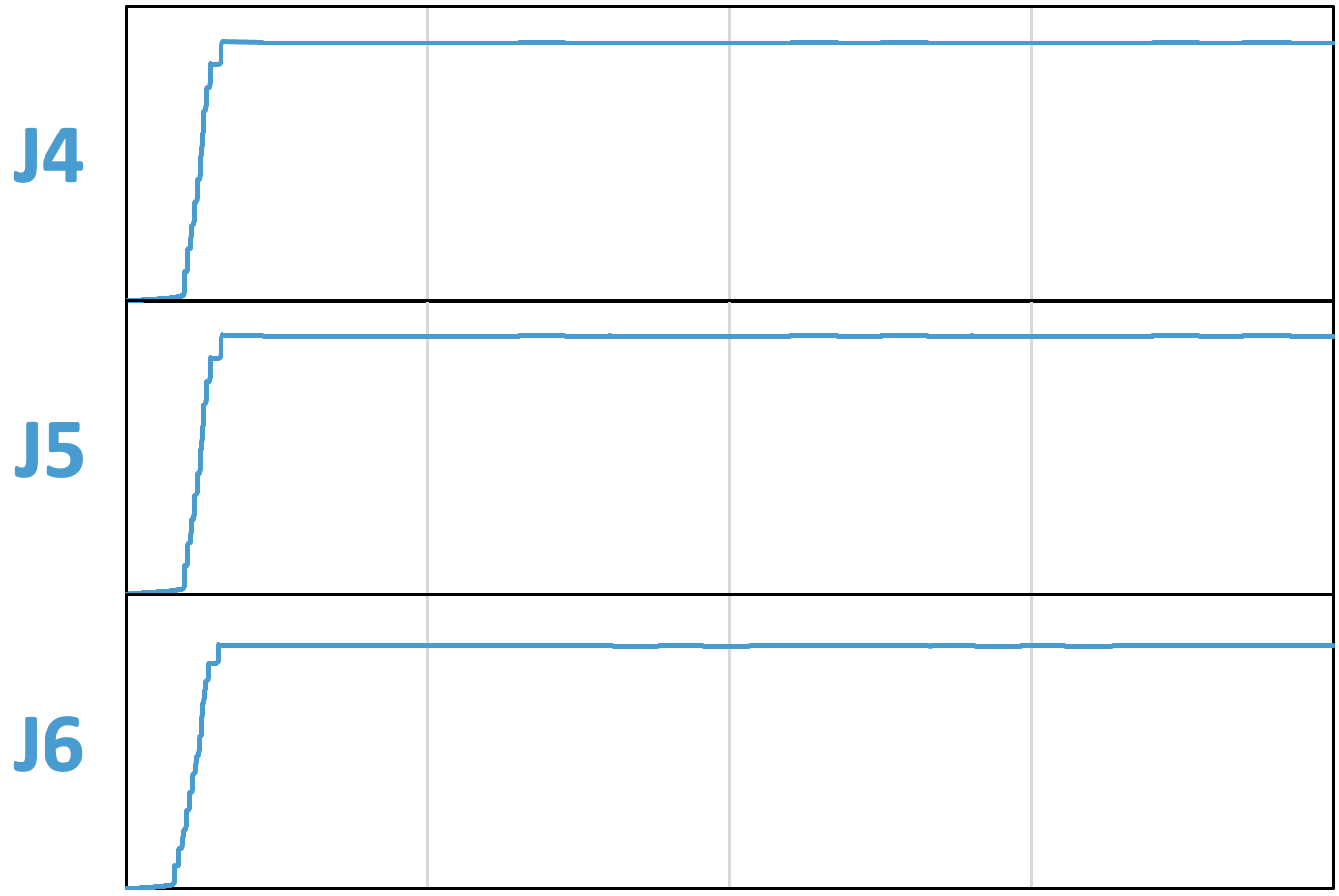}
    }\\
    \hspace{0.01cm}
    \subfloat[\label{subfig:celem31}]{%
    \includegraphics[width=0.35\textwidth]{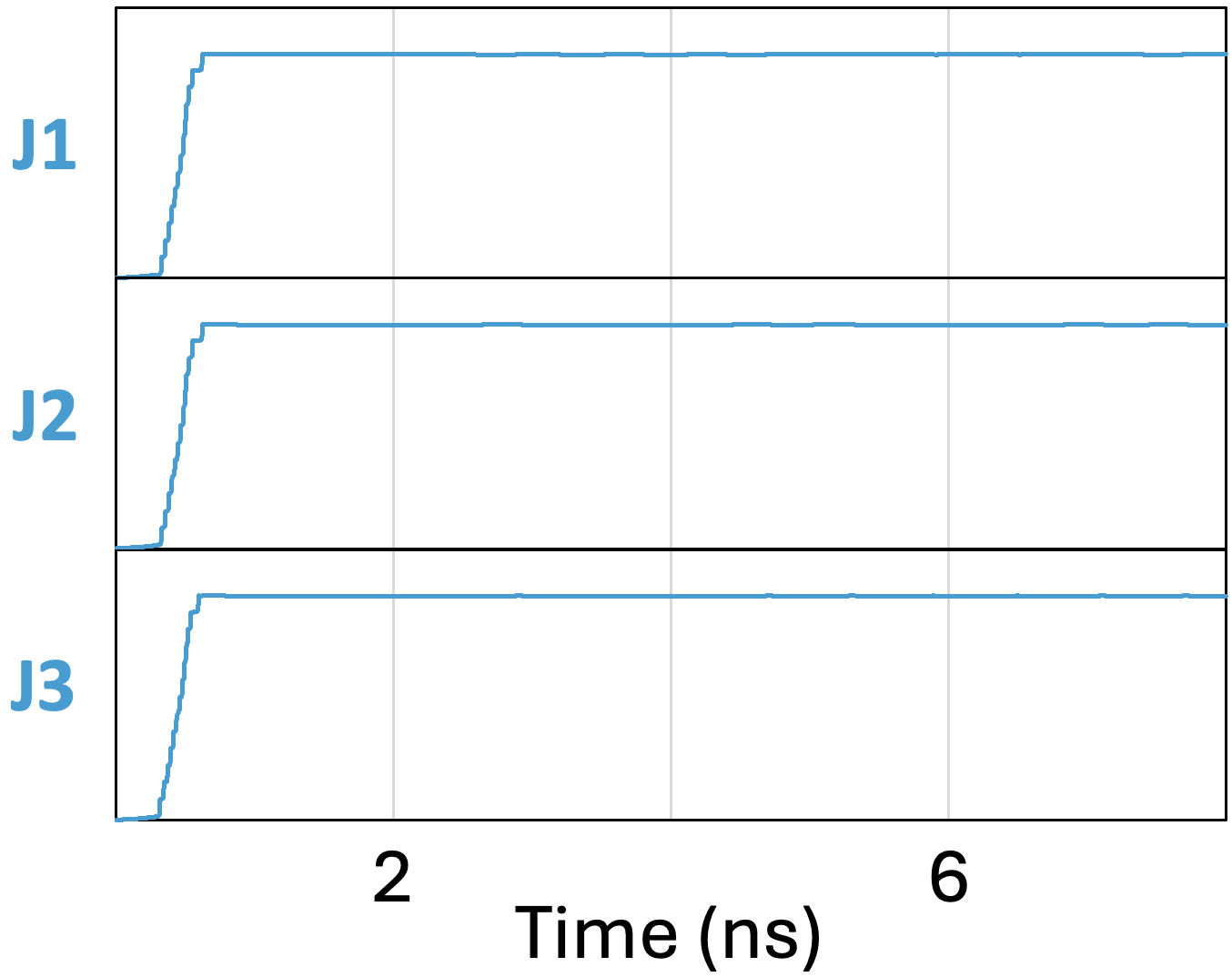}
    }
    \hspace{0.75cm}
    \subfloat[\label{subfig:icelem41}]{%
      \includegraphics[width=0.35\textwidth]{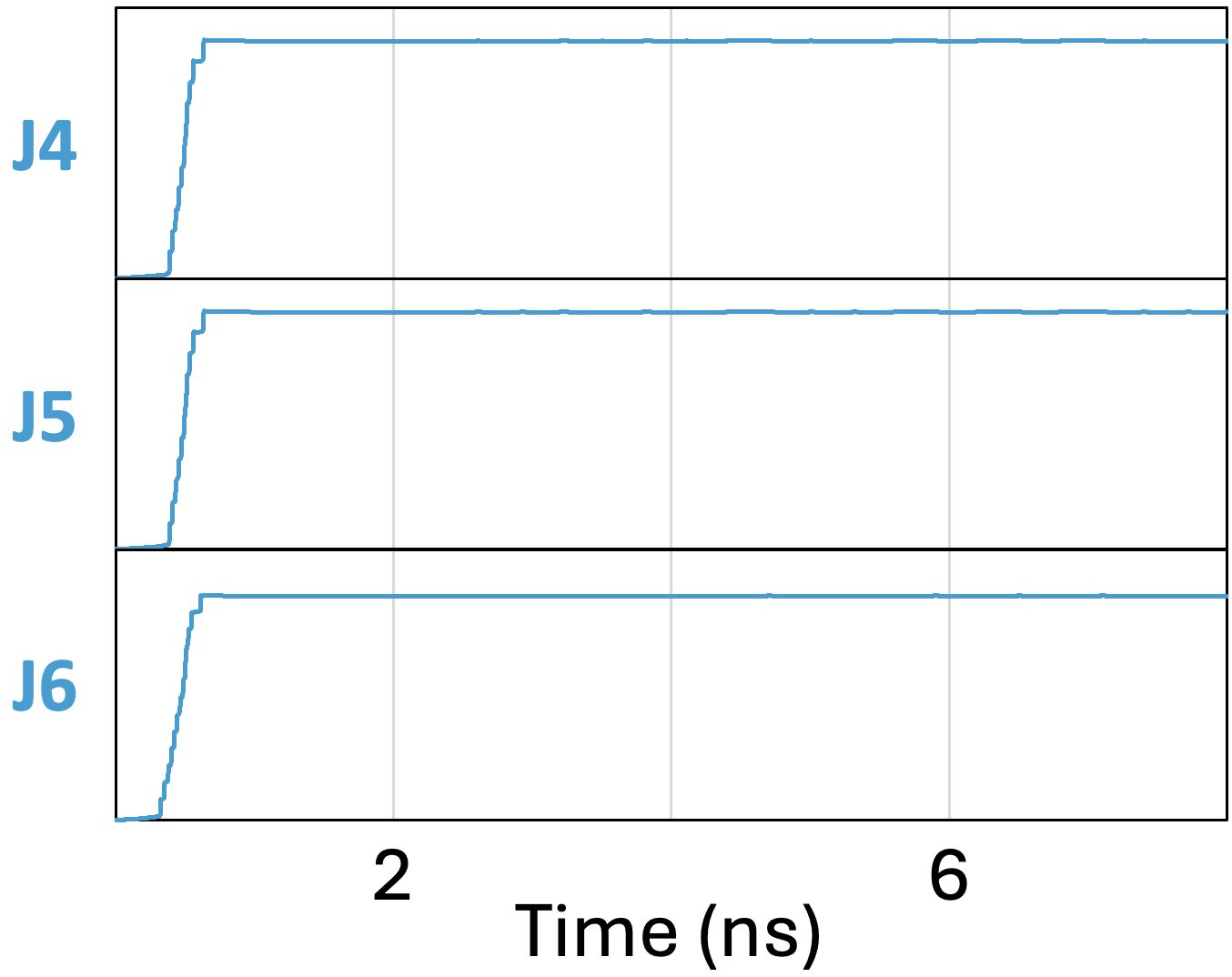}
    }\\
    \caption{Panel a: Simulation waveform showing the phases of the bias JJs in the C-element labeled 1 in the full adder depicted in Figure~\ref{fig:fulladder_block} for the input sequences shown in Figure~\ref{fig:fulladder_wvfm}. Panel b: Bias JJ phases in the Inverted C-element labeled 2 in the same full adder. Panel c: Bias JJ phases in the C-element labeled 3. Panel d: Bias JJ phases in the Inverted C-element labeled 4. In all cases, zero phase imbalance is demonstrated.}
    \label{fig:enc}
  \end{figure}

\subsection{xeSFQ gate composability}
To confirm that our hypothesis regarding zero static power consumption extends beyond a single xeSFQ gate, we simulated a dual-rail 1-bit full adder. The schematic of this design, shown in Figure~\ref{fig:fulladder_block}, is adopted from the original xSFQ publication~\cite{xsfq}. All \textsc{AND} gates are implemented as C-elements (Figure~\ref{fig:mealy}a), and all \textsc{OR} gates as Inverted C-elements (Figure~\ref{fig:mealy}b). An example input pattern demonstrates how the logic gates process the data as it propagates through the design. 

Figure~\ref{fig:fulladder_wvfm} demonstrates correct functionality across all possible input combinations. Figure~\ref{fig:enc} shows the phases of the bias JJs for the gates highlighted in Figure~\ref{fig:fulladder_block}, with the JJ labeling consistent with that introduced in Figure~\ref{fig:schema}. As observed in the single-gate xeSFQ circuit, the phase waveforms remain flat, indicating no phase accumulation and confirming zero static power consumption after the biasing startup sequence. For comparison, a 1-bit ERSFQ full adder, composed of eight \textsc{AND}, \textsc{OR}, \textsc{XOR}, and \textsc{DFF} gates, can have a static power consumption of up to 72\% that of the dynamic power with arbitrary input sequences. This exceeds the percentage found at the gate level analysis because the increased number of datapaths create a higher rate of phase imbalance.




\subsection{Benchmarking}
To evaluate the resource and energy requirements of the proposed xeSFQ logic family, we use the ISCAS85~\cite{iscas85} and EPFL~\cite{epfl} benchmark circuits. The primary metric for resource utilization is the total number of JJs. The results are based on synthesized xSFQ designs~\cite{volk2024synthesis} and account for all associated overheads, including fan-out and passive transmission line-based interconnections between gates. Table~\ref{tab:bench} summarizes our synthesis results. For dynamic energy calculations, the switching activity in all designs is set to one~\footnote{Despite exhibiting higher switching activity compared to traditional SFQ approaches based on direct encoding, prior research~\cite{xsfq} has demonstrated that the dynamic energy efficiency of x(e)SFQ is significantly lessened compared to its counterparts---by more than one order of magnitude---thanks to optimized pipelining and a lower JJ count for the same functionality.}. Notably, all designs are combinational, do not require a clock signal, and are executed in a single cycle.

\begin{table}[]
\caption{Synthesis results for ISCAS85 and EPFL benchmark circuits. C- and Inverted C-element-related parameters are denoted by \textsc{C} and \textsc{IC}, respectively. The energy numbers used for individual cells are as follows: \textsc{C}=2.3~aJ, \textsc{IC}=2.3~aJ, splitter=1.8~aJ, PTL driver=1.0~pJ, PTL receiver=1.3~aJ. Regarding JJ counts, \textsc{C}=4~JJs (+3 bias JJs), \textsc{IC}=4~JJs (+3 bias JJs), splitter=3~JJs (+1 bias JJ), PTL driver=2~JJs (+1 bias JJ), PTL receiver=3~JJs (+2 bias JJs).}
\centering
\resizebox{0.9\textwidth}{!}{%
\begin{tabular}{|c|c|c|c|c|c|c|c|}
\hline
\textbf{Circuit} & \textbf{\begin{tabular}[c]{@{}c@{}}\#C/IC \\ cells\end{tabular}} & \textbf{\begin{tabular}[c]{@{}c@{}}\#C/IC \\ JJs\end{tabular}} & \textbf{\begin{tabular}[c]{@{}c@{}}\#fan-out \\ JJs\end{tabular}} & \textbf{\begin{tabular}[c]{@{}c@{}}\#PTL \\ JJs\end{tabular}} & \textbf{\begin{tabular}[c]{@{}c@{}}\#bias \\ JJs\end{tabular}} & \textbf{\begin{tabular}[c]{@{}c@{}}Dynamic \\ energy (fJ)\end{tabular}} & \textbf{\begin{tabular}[c]{@{}c@{}}Static \\ energy (fJ)\end{tabular}} \\ \hline
c880             & 452                                                               & 1,808                                                           & 1,134                                                             & 4,520                                                         & 4,446                                                          & 3.8                                                                    & 0                                                                      \\ \hline
c1908            & 503                                                               & 2,012                                                           & 1,386                                                             & 5,030                                                         & 4,989                                                          & 4.3                                                                    & 0                                                                      \\ \hline
c499             & 682                                                               & 2,728                                                           & 1,896                                                             & 6,820                                                         & 6,770                                                          & 5.9                                                                   & 0                                                                      \\ \hline
c3540            & 1,646                                                             & 6,584                                                           & 4,704                                                             & 16,460                                                        & 16,382                                                         & 14.3                                                                   & 0                                                                      \\ \hline
c5312            & 1,944                                                             & 7,776                                                           & 5,421                                                             & 19,440                                                        & 19,303                                                         & 16.8                                                                   & 0                                                                      \\ \hline
c7752            & 2,571                                                             & 10,284                                                          & 6,873                                                             & 25,710                                                        & 25,430                                                         & 22.0                                                                   & 0                                                                      \\ \hline
int2float        & 225                                                               & 900                                                             & 630                                                               & 2,250                                                         & 2,235                                                          & 1.9                                                                    & 0                                                                      \\ \hline
dec              & 304                                                               & 1,216                                                           & 1,632                                                               & 3,040                                                         & 3,280                                                          & 3.1                                                                    & 0                                                                      \\ \hline
priority         & 892                                                               & 3,568                                                           & 1,935                                                             & 8,920                                                         & 8,673                                                          & 7.4                                                                   & 0                                                                      \\ \hline
sin              & 9,977                                                             & 39,908                                                          & 29,862                                                            & 99,770                                                        & 99,747                                                         & 87.2                                                                 & 0                                                                      \\ \hline
cavlc            & 721                                                               & 2,884                                                           & 2,136                                                             & 7,210                                                         & 7,201                                                          & 6.3                                                                   & 0                                                                      \\ \hline
\end{tabular}
}
\label{tab:bench}
\end{table}

\label{sec:eval}

\section{Conclusion}
Superconductor electronics are still largely dominated by SFQ logic families, with many projections for power-constrained applications pointing to ERSFQ as the next-generation solution, primarily due to its theoretical zero static power. While ERSFQ is indeed more energy-efficient than RSFQ---due to the elimination of the resistive bias network---the presented research reveals that in realistic scenarios, its static power consumption is not zero and can, in some cases, be on par with that of dynamic power. A closer examination attributes this behavior to phase accumulation, a consequence of direct encoding, in which the presence or absence of an SFQ pulse directly represents the encoded value. To address this, we introduce xeSFQ, which integrates xSFQ’s alternating, balanced encoding and logic abstractions with ERSFQ’s biasing scheme. The result is simple, compact, clock-free SFQ circuit designs with truly zero static power consumption. This work not only boosts energy efficiency for SFQ logic implementations but also provides deeper insights into the bias-logic relationship, demonstrating that data representation can impact both dynamic and static power consumption.

\section*{References}
\bibliography{sections/refs.bib}

\begin{thebibliography}{10}

\bibitem{superrl}
G.~Tzimpragos, D.~Vasudevan, N.~Tsiskaridze, G.~Michelogiannakis, A.~Madhavan, J.~Volk, J.~Shalf, and T.~Sherwood, ``A computational temporal logic for superconducting accelerators,'' in {\em Proceedings of the Twenty-Fifth International Conference on Architectural Support for Programming Languages and Operating Systems}, ASPLOS '20, (New York, NY, USA), p.~435–448, Association for Computing Machinery, 2020.

\bibitem{largescale}
D.~S. Holmes, A.~M. Kadin, and M.~W. Johnson, ``Superconducting computing in large-scale hybrid systems,'' {\em Computer}, vol.~48, no.~12, pp.~34--42, 2015.

\bibitem{770ghz}
W.~Chen, A.~Rylyakov, V.~Patel, J.~Lukens, and K.~Likharev, ``Rapid single flux quantum t-flip flop operating up to 770 ghz,'' {\em IEEE Transactions on Applied Superconductivity}, vol.~9, no.~2, pp.~3212--3215, 1999.

\bibitem{radebaugh2022review}
R.~Radebaugh, ``Review of refrigeration methods,'' in {\em Handbook of Superconductivity}, pp.~501--518, CRC Press, 2022.

\bibitem{histaticpwr}
S.~Polonsky, ``Delay insensitive rsfq circuits with zero static power dissipation,'' {\em IEEE Transactions on Applied Superconductivity}, vol.~9, no.~2, pp.~3535--3538, 1999.

\bibitem{rsfq}
K.~K. Likharev and V.~K. Semenov, ``Rsfq logic/memory family: A new josephson-junction technology for sub-terahertz-clock-frequency digital systems,'' {\em IEEE Transactions on Applied Superconductivity}, vol.~1, no.~1, pp.~3--28, 1991.

\bibitem{ersfqesfq}
O.~A. Mukhanov, ``Energy-efficient single flux quantum technology,'' {\em IEEE Transactions on Applied Superconductivity}, vol.~21, no.~3, pp.~760--769, 2011.

\bibitem{ersfq_currentredistribution}
M.~B. Ketchen, J.~Timmerwilke, G.~Gibson, and M.~Bhushan, ``Ersfq power delivery: A self-consistent model/hardware case study,'' {\em IEEE Transactions on Applied Superconductivity}, vol.~29, no.~7, pp.~1--11, 2019.

\bibitem{switchingBiasJJs_ERSFQ}
I.~Vernik, S.~Kaplan, M.~Volkmann, A.~Dotsenko, C.~Fourie, and O.~Mukhanov, ``Design and test of asynchronous esfq circuits,'' {\em Superconductor Science and Technology}, vol.~27, no.~4, p.~044030, 2014.

\bibitem{supertools}
G.~Pasandi, {\em Designing Efficient Algorithms and Developing Suitable Software Tools to Support Logic Synthesis of Superconducting Single Flux Quantum Circuits}.
\newblock PhD thesis, University of Southern California, 2020.

\bibitem{fakeAsync}
H.~Gerber, C.~Fourie, and W.~Perold, ``Optimised asynchronous self-timing for superconducting rsfq logic circuits,'' in {\em 2004 IEEE Africon. 7th Africon Conference in Africa (IEEE Cat. No.04CH37590)}, vol.~1, pp.~551--556 Vol.1, 2004.

\bibitem{dffinsertion}
N.~Kito, K.~Takagi, and N.~Takagi, ``Logic-depth-aware technology mapping method for rsfq logic circuits with special rsfq gates,'' {\em IEEE Transactions on Applied Superconductivity}, vol.~32, no.~4, pp.~1--5, 2022.

\bibitem{optpipeline}
A.~Hartstein and T.~Puzak, ``Optimum power/performance pipeline depth,'' in {\em Proceedings. 36th Annual IEEE/ACM International Symposium on Microarchitecture, 2003. MICRO-36.}, pp.~117--125, 2003.

\bibitem{jimpipeline}
S.~R. Kunkel and J.~E. Smith, ``Optimal pipelining in supercomputers,'' in {\em Proceedings of the 13th Annual International Symposium on Computer Architecture}, ISCA '86, (Washington, DC, USA), p.~404–411, IEEE Computer Society Press, 1986.

\bibitem{xsfq}
G.~Tzimpragos, J.~Volk, A.~Wynn, J.~E. Smith, and T.~Sherwood, ``Superconducting computing with alternating logic elements,'' in {\em 2021 ACM/IEEE 48th Annual International Symposium on Computer Architecture (ISCA)}, pp.~651--664, 2021.

\bibitem{dsfq}
S.~V. Rylov, ``Clockless dynamic sfq and gate with high input skew tolerance,'' {\em IEEE Transactions on Applied Superconductivity}, vol.~29, no.~5, pp.~1--5, 2019.

\bibitem{pcl}
Q.~Herr, T.~Josephsen, and A.~Herr, ``{Superconducting pulse conserving logic and Josephson-SRAM},'' {\em Applied Physics Letters}, vol.~122, p.~182604, 05 2023.

\bibitem{rql}
Q.~P. Herr, A.~Y. Herr, O.~T. Oberg, and A.~G. Ioannidis, ``{Ultra-low-power superconductor logic},'' {\em Journal of Applied Physics}, vol.~109, p.~103903, 05 2011.

\bibitem{volk2024synthesis}
J.~Volk, P.~Papanikolaou, G.~Zervakis, and G.~Tzimpragos, ``Synthesis of resource-efficient superconducting circuits with clock-free alternating logic,'' {\em arXiv preprint arXiv:2407.20942}, 2024.

\bibitem{iscas85}
M.~Hansen, H.~Yalcin, and J.~Hayes, ``Unveiling the iscas-85 benchmarks: a case study in reverse engineering,'' {\em IEEE Design \& Test of Computers}, vol.~16, no.~3, pp.~72--80, 1999.

\bibitem{epfl}
L.~Amarù, P.-E. Gaillardon, and G.~De~Micheli, ``The epfl combinational benchmark suite,'' 2015.

\bibitem{biasind200pH}
A.~F. Kirichenko, I.~V. Vernik, O.~A. Mukhanov, and T.~A. Ohki, ``Ersfq 4-to-16 decoder for energy-efficient ram,'' {\em IEEE Transactions on Applied Superconductivity}, vol.~25, no.~3, pp.~1--4, 2015.

\bibitem{phasedef}
J.~A. Blackburn, M.~Cirillo, and N.~Grønbech-Jensen, ``A survey of classical and quantum interpretations of experiments on josephson junctions at very low temperatures,'' {\em Physics Reports}, vol.~611, pp.~1--33, 2016.

\bibitem{tinkham}
T.~Michael, {\em Introduction to Superconductivity (2nd Edition)}.
\newblock Dover Publications, 1996.

\bibitem{wrspice}
W.~R. Incorporated, ``{WRspice} reference manual,'' tech. rep., June 2019.

\bibitem{sfq5ee}
S.~K. Tolpygo, V.~Bolkhovsky, T.~J. Weir, A.~Wynn, D.~E. Oates, L.~M. Johnson, and M.~A. Gouker, ``Advanced fabrication processes for superconducting very large-scale integrated circuits,'' {\em IEEE Transactions on Applied Superconductivity}, vol.~26, no.~3, pp.~1--10, 2016.

\bibitem{fjtl}
M.~E. Çelik, T.~V. Filippov, D.~Kirichenko, M.~Habib, A.~Talalaevskii, A.~Sahu, and D.~Gupta, ``25-ghz operation of ersfq time-to-digital converter,'' {\em IEEE Transactions on Applied Superconductivity}, vol.~31, no.~5, pp.~1--5, 2021.

\bibitem{fjtl_margins}
N.~K. Katam, O.~Mukhanov, and M.~Pedram, ``Simulation analysis and energy-saving techniques for ersfq circuits,'' {\em IEEE Transactions on Applied Superconductivity}, vol.~29, no.~5, pp.~1--7, 2019.

\end{thebibliography}

\end{document}